\documentclass{emulateapj}
\usepackage{apjfonts}
\slugcomment{Draft Version}

\begin{document}


\title{Evidence for nuclear radio jet and its structure down to $\lesssim$100
Schwarzschild radii \\ in the center of the Sombrero Galaxy (M~104, NGC~4594)}

\author{
Kazuhiro Hada\altaffilmark{1,2}, 
Akihiro Doi\altaffilmark{3},
Hiroshi Nagai\altaffilmark{2}, 
Makoto Inoue\altaffilmark{4}, \\
Mareki Honma\altaffilmark{2,5},
Marcello Giroletti\altaffilmark{1}, and 
Gabriele Giovannini\altaffilmark{1,6}
}

\affil{$^1$INAF Istituto di Radioastronomia, via Gobetti 101, I-40129
Bologna, Italy}

\affil{$^2$National Astronomical Observatory of Japan,
Osawa, Mitaka, Tokyo 181-8588, Japan} 

\affil{$^3$Institute of Space and
Astronautical Science, Japan Aerospace Exploration Agency, 3-1-1
Yoshinodai, Chuo, Sagamihara 252-5210, Japan} 

\affil{$^4$Academia Sinica Institute for Astronomy and Astrophysics, 11F
Astronomy-Mathematics Bldg., National Taiwan University, No. 1, Roosevelt Rd.,
Sec. 4 Taipei 10617, Taiwan R.O.C.}

\affil{$^5$Department of Astronomical Science, The Graduate University
for Advanced Studies (SOKENDAI), 2-21-1 Osawa, Mitaka, Tokyo 181-8588,
Japan}

\affil{$^6$Dipartimento di Fisica e Astronomia, Universit\`a di Bologna, via
Ranzani 1, I-40127 Bologna, Italy}

\begin{abstract}
Sombrero galaxy (M~104, NGC~4594) is associated with one of the nearest
low-luminosity active galactic nuclei (LLAGN). We investigated the detailed radio
structure of the nucleus of the Sombrero using high-resolution,
quasi-simultaneous, multi-frequency, phase-referencing Very-Long-Baseline-Array
observations. We obtained the VLBI images toward this nucleus, with unprecedented
sensitivities and resolutions, at the seven frequencies between 1.4 and 43~GHz, in
which those at 15, 24 and 43~GHz are the first clear VLBI detections and imaging
for this source. At 43~GHz, where the highest resolution is available, the nuclear
structure was imaged on a linear scale under 0.01~pc or 100 Schwarzschild radii,
revealing the compact radio core with a high brightness temperature of $\gtrsim 3
\times 10^9$~K. For the first time, we have discovered the presence of the
extended structure in this nucleus, which is directing from the radio core in two
sides toward northwest/southeast directions. The nuclear structure shows a clear
spatial gradient on the radio spectra; the core region is mildly inverted whereas
the extended region becomes progressively steep, as commonly seen in more luminous
AGN with powerful relativistic jets. Moreover, the radio core shows a
frequency-dependent size with an elongated shape toward the same direction of the
extended structure, and the position of the radio core also tends to be frequency
dependent. A set of these new findings provide evidence that the central engine of
the Sombrero is powering radio jets, and the jets are overwhelming the emission
from the underlying radiatively-inefficient accretion flow over the observed
frequency range. Based on the observed brightness ratio of jet-to-counter jet,
core position shift and its comparision with a theoretical model, we constrained
the following fundamental physical parameters for the M~104 jets: (1) the northern
side is the approaching jet, whereas the southern side is receding: (2) the
inclination angle of the jet is relatively close to our line-of-sight, probably
less than $\sim$$25^{\circ}$: (3) the jet intrinsic velocity is highly
sub-relativistic at a speed less than $\sim$$0.2~c$.  The derived pole-on nature of
the M~104 jet is in accordance with the previous argument that M~104 contains a
true type II AGN, i.e., the broad line region of this nucleus is actually absent
or intrinsically weak, if the plane of the presumed circumnuclear torus is
perpendicular to the axis of the radio jets.
\end{abstract}


\section{Introduction}
M~104 (NGC~4594) is a famous early-type spiral galaxy known as the
``Sombrero''. It is located in the southern hemisphere at a distance of
$D=9.0\pm0.1$~Mpc~\citep{spitler2006} and the galactic plane shows a nearly
edge-on view~\citep[$84^{\circ}$ from our
line-of-sight;][]{rubin1985}. \textit{Hubble Space Telescope} observations
indicate the presence of a supermassive black hole weighing
$1\times10^9~M_{\odot}$ at the center~\citep{kormendy1988, kormendy1996}. The
nucleus is classified into a low-ionization nuclear emission region (LINER) based
on its optical emission-line property~\citep{heckman1980}, belonging to a
representative sub-class of low-luminosity active galactic nuclei (LLAGN).

\begin{table*}[ttt]
 \begin{minipage}[t]{1.0\textwidth}
  \centering 
  \caption{VLBA observations of M~104} \smallskip
    \begin{tabular*}{1.0\textwidth}{@{\extracolsep{\fill}}lccccc}
    \hline
    \hline
    UT Date  & Frequency & $\Delta \nu$ & Beam size & $I_{\rm p}$    & $I_{\rm
     rms}$ \\
          & (GHz) &  (MHz)  & (mas$\times$mas, deg.) & (mJy/bm) & (mJy/bm) \\
          &      & (a)     & (b)     & (c)     & (d)    \\
    \hline
    2008 Apr 26$^{\dag}$................. & 1.430 & 32 & 23.6 $\times$ 5.70, $-17$  & 51.8  & 0.068 \\
    2011 Mar 23..................         & 2.266 & 32 & 9.44 $\times$ 3.83, $-4.2$ & 52.2  & 0.139 \\
                                          & 4.990 & 64 & 4.53 $\times$ 1.95, $9.3$  & 67.6  & 0.091 \\
                                          & 8.416 & 32 & 2.42 $\times$ 1.03, $1.1$  & 66.7  & 0.101 \\
                                          & 15.36 & 64 & 1.43 $\times$ 0.54, $-6.3$ & 72.9  & 0.215 \\
                                          & 23.80 & 64 & 0.97 $\times$ 0.37, $-8.7$ & 77.9  & 0.215 \\
                                          & 43.21 & 64 & 0.86 $\times$ 0.23, $-17$  & 82.2  & 0.51 \\
    2011 Mar 30..................         & 2.266 & 32 & 9.75 $\times$ 4.08, $-6.8$ & 53.7  & 0.149 \\
                                          & 4.990 & 64 & 3.64 $\times$ 1.55, $-2.8$ & 64.8  & 0.088 \\
                                          & 8.416 & 32 & 2.47 $\times$ 1.01, $-3.3$ & 68.4  & 0.111 \\
                                          & 15.36 & 64 & 1.41 $\times$ 0.54, $-4.5$ & 77.5  & 0.167 \\
                                          & 23.80 & 64 & 1.01 $\times$ 0.35, $-8.4$ & 81.0  & 0.257 \\
                                          & 43.21 & 64 & 0.73 $\times$ 0.24, $-12.4$ & 87.9 & 1.03 \\
    \hline
    \end{tabular*} \smallskip
  \end{minipage}
  \label{tab:img_prm} Notes: $\dag$ VLBA archival data (without Saint Croix, no
 phase-referencing mode): (a) total bandwidth: (b) synthesized beam with
 naturally-weighted scheme: (c) peak intensity of self-calibrated images of M~104
 under naturally-weighting scheme (see Figure~1): (d) rms image noise level of
 M~104 images under naturally-weighting scheme.
\end{table*}

Physical processes acting in the vicinity of the LLAGN central engines remain as a
major question in the local Universe~\citep[][and the references therein]{ho1997a,
ho2008}. Because of the absence of the big blue bump and highly sub-Eddington
luminosity in their spectral energy distributions (SED), the accretion state is
thought to be described by advection-dominated accretion flow (ADAF) or
radiatively-inefficeint accretion flow (RIAF), which is geometrically-thick and
optically-thin with a hot electron temperature of
$\sim$$10^{9}$~K~\citep{narayan1994, narayan1995}. Bremsstrahlung and inverse
Compton emission from such flow reasonably explain the above characteristics of
LLAGN SED as well as observed harder X-ray spectral
nature~\citep[e.g.,][]{manmoto1997, narayan1998, quataert1999b}.  In contrast,
observed LLAGN spectra at radio bands are typically flat or mildly inverted
\citep[$\alpha < 0.2$\footnote{We define spectral index as $S_{\nu} \propto
\nu^{+\alpha}$ in the present paper.};][]{nagar2001, doi2005, doi2011} compared to
that expected by thermal synchrotron from RIAF~\citep[$\alpha \gtrsim
0.4$;][]{mahadevan1997}.  This may be reconciled by including nonthermal electron
populations in RIAF~\citep{yuan2003, liu2013} or considering convection
processes~\citep{narayan2000}. Alternatively, as a simple analogy of more luminous
AGN, such a discrepancy can be naturally accommodated if outflowing radio jets are
present~\citep[][\maketitle hereafter FB99]{falcke1995, falcke1999}

Given their structural compactness and the radio regime being a key band,
Very-Long-Baseline-Interferometory (VLBI) is a powerful tool to understand the
nature of LLAGN activities. The nuclear structures are often unresolved even on
milliarcsecond (mas) scale, constraining the size of radio emitting region to be
$\lesssim 10^{3-4}$ Schwarzschild radii $(R_{\rm s})$ with a brightness
temperature of $\gtrsim 10^{8-9}$~K~\citep{ulvestad2001}. These unresolved cores
show flat or only mildly inverted radio spectra, and their luminosities are
generally too high to account for RIAF~\citep{anderson2004}. LLAGN with higher
radio core luminosities tend to show elongated or extended structures on pc/sub-pc
scale~\citep{falcke2000, nagar2002, nagar2005, giroletti2005}. Especially, in a
couple of the best studied nearby cases such as M~81 and NGC~4258, the nuclear
radio jets are evidently resolved together with an offset of the radio emitting
site from their putative dynamical centers~\citep{bietenholz2000, bietenholz2004,
herrnstein1997, marti2011}, implying that significant amount of their accretion
energy is channeled into outflowing jets~\citep{markoff2008, yuan2002, doi2013}.

Along with these sources, the nucleus of M~104 also represents an ideal case. A
particularly striking advantange is its accessibility to the immediate vicinity of
the central engine; due to its proximity and a large estimated black hole mass,
their combination yields $0.1~{\rm mas} = 0.004~{\rm pc} = 45.4~R_{\rm s}$. Such a
physical scale in terms of $R_{\rm s}$ is $\sim$5 or $\sim$20 times finer than
that of M~81 or NGC~4258, and even comparable to the accessible scales for the
Galactic Center SgrA* or the nearby radio galaxies M~87 and
Cen~A~\citep{doi2009}\footnote{In terms of the black hole mass of Cen~A,
\citet{doi2009} assume $\sim$$2 \times
10^8~M_{\odot}$~\citep{marconi2001}. However, more recent measurements revise its
mass toward a smaller value $\sim$$5 \times 10^7~M_{\rm \odot}$~\citep[see][and
references therein]{neumayer2010}. In this case, M~104 has a larger apparent
diameter of the central black hole than that of Cen~A.}.

The M~104 nucleus has been intensively studied over the wide range of
electromagnetic spectrum. At radio frequencies, the nucleus is completely
pointlike on sub-arcsecond scale~\citep{bruyn1976, krause2006}. Non-simultaneous
radio spectrum on arcsec scale is roughly flat with a typical flux density of
$\sim$100~mJy, but variabilities are reported on some of different amplitudes and
time scales~\citep{bruyn1976, hummel1984, bajaja1988}. Recent JCMT observations
indicate the dominance of the AGN emission up to 350~GHz~\citep{bendo2006}. X-ray
observations at 2-10 keV band show a pointlike, hard ($\Gamma=1.89$) spectral
shape nucleus with the absence of Fe~K lines~\citep{pellegrini2003}. Bolometric
luminosity of the nucleus is highly sub-Eddington~\citep[$L_{\rm
bol}\sim2.5\times10^{41}~{\rm erg~s^{-1}} \sim 2\times 10^{-6}L_{\rm
Edd}$;][]{pellegrini2003}, and the overall characteristics of its broadband SED is
similar to those of other LLAGN~\citep{ho1999}.

Despite such observational advantages, little is known yet about its
milliarcsecond-scale structure. Early VLBI experiments of this nucleus shows a
single unresolved component ($<2.7$~mas) without any signs of extended
feature~\citep{graham1981, preston1985, shaffer1979}. More recently, a few of
snap-shot images are obtained from geodetic RDV\footnote{Research and Development
with the Very-Long-Baseline-Array (VLBA); e.g., \citet{gordon2005}.} observations
at 2 and 8~GHz, but the nucleus still looks point like. However, the sensitivities
and the resolutions of these observations are relatively poor; to uncover the
exact nature of its nuclear structure, more dedicated observations are
indispensable. Indeed, some of theoretical studies predict a structual complexity
for the nuclear structure of M~104~\citep{dimatteo2001, yuan2009}.

In this paper, we report on our deep VLBI observations toward the nucleus of
M~104. Thanks to the use of quasi-simultaneous, multi-frequency, phase-referencing
mode at a higher sensitivity, we have obtained a number of new findings on the
nuclear structure of this source. The data and the analysis are described in the
next section. We show the results in section 3, and then discuss the physical
properties of the M~104 nucleus in section~4. Finally, we summarize our main
conclusions in section~5.

\begin{figure*}[ttt]
 \centering \includegraphics[angle=0,width=1.0\textwidth]{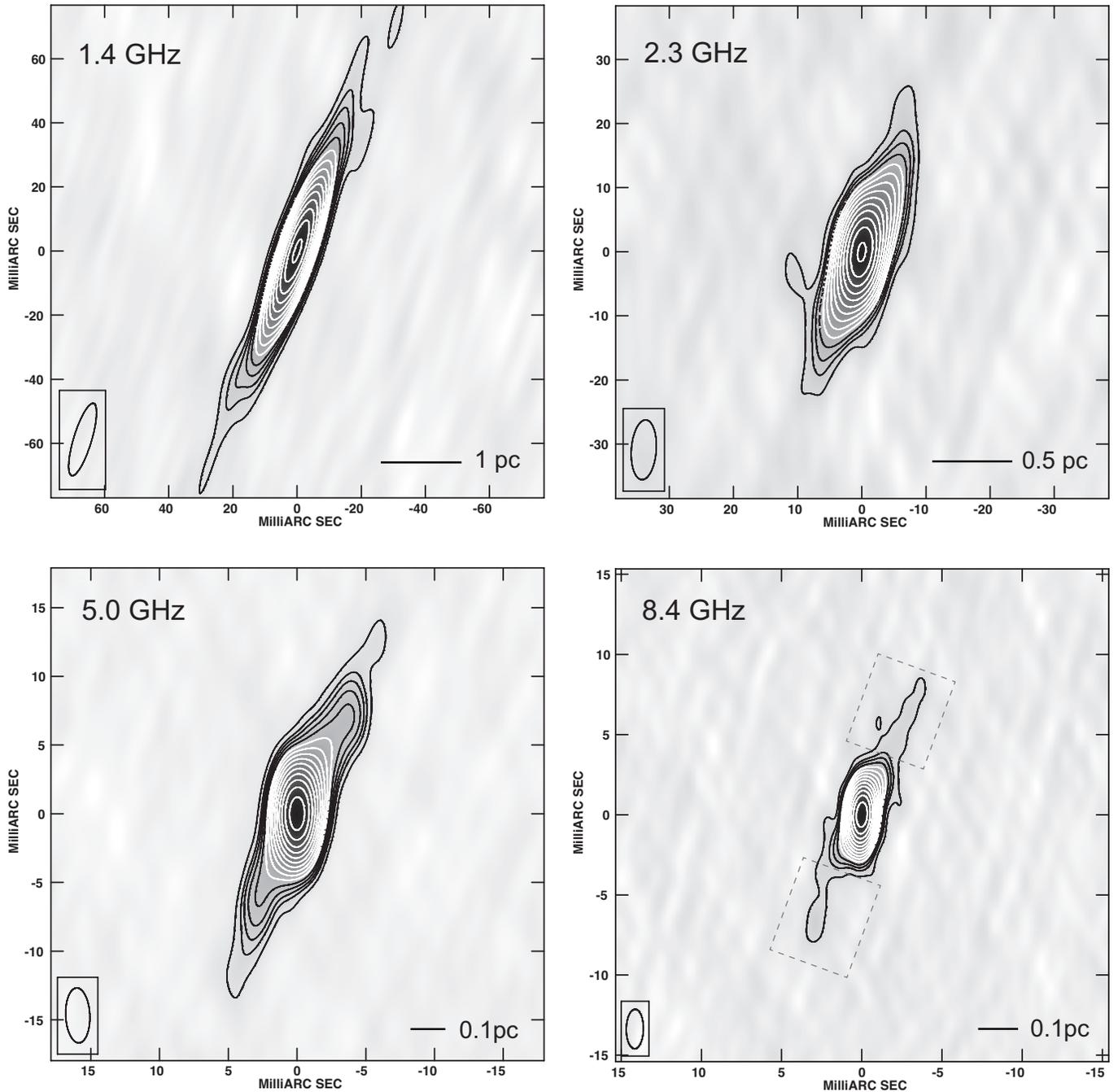}
 \caption{VLBA images of the M~104 nucleus at all the observed
 frequencies. Naturally-weighted scheme is used for each data and the synthsized
 beam are shown in bottom-left corner of each image. Contours start from $-1$, 1,
 2, ... times 3$\sigma$ image rms level and increasing by factors of
 $2^{1/2}$. The length of the horitontal bar in bottom-right corner of each image
 indicates a linear scale of each image on pc unit. Two rectangular areas enclosed
 by dashed lines on 8.4~GHz image ($6\times 5$~mas$^{2}$ in $P.A.=-20^{\circ}$ or
 $P.A.=160^{\circ}$ starting from 4~mas distance from the core peak for each
 rectangle) indicate the regions where the brightness ratio of the
 northern/southern extended components are measured by integrating the CLEAN
 components (adopted rectangular regions at each frequency are summarized in
 Table~2).}  \label{fig:images}
\end{figure*}

\setcounter{figure}{0}

\begin{figure*}[ttt]
 \centering
 \includegraphics[angle=0,width=1.0\textwidth]{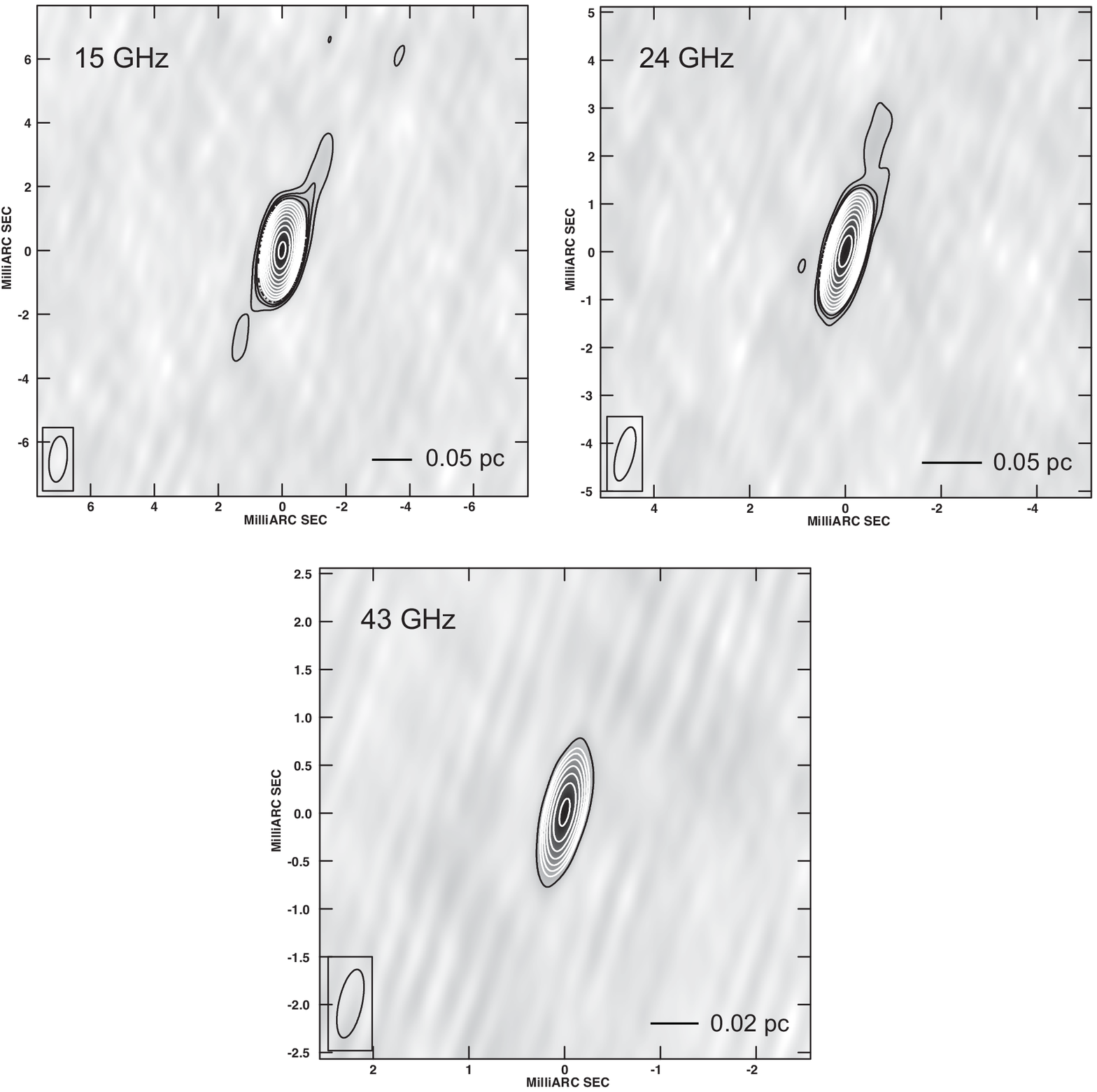}
 \caption{Continued.}
 \label{fig:}
\end{figure*}

\section{Observations and data reduction}
On 23 and 30 March 2011, we observed the M~104 nucleus with VLBA at 2.3, 5.0, 8.4,
15.4, 23.8 and 43.2~GHz quasi-simultaneously. The observations involved rapid
switching between M~104 and the nearby radio source J1239--1023, which is
separated by 1.\hspace{-.3em}$^{\circ}$23 on the sky, plus a few scans of the
bright fringe-finder source 3C~279. To reduce systematic errors and to achieve
nearly the same $uv$ coverage among frequencies, we carried out observations by
alternating each frequency in turn every 10$\sim$40 minutes. Each session was
conducted in a 10-hour overnight run at a data recording speed of 512 Mega-bit per
second. All of the 10 VLBA stations were participated.

The initial data calibration was performed using the National Radio Astronomy
Observatory (NRAO) Astronomical Image Processing System (AIPS) based on the
standard VLBI data reduction procedures. The amplitude calibration with opacity
corrections was applied using the measured system noise temperature and the
elevation-gain curve of each antenna. We next performed apriori corrections for
the visibility phases; antenna parallactic angle differences between M~104 and
J1239--1023, ionospheric dispersive delays using the ionospheric model provided by
the Jet Propulsion Laboratory (JPL), and instrumental delays/phases using a scan
of 3C~279 were corrected. Then, to create images of M~104 and J1239--1029, we
performed a fringe-fitting on each source separately and removed residual delays,
rates and phases assuming a point source model.  We clearly detected fringes for
these data with sufficient signal-to-noise ratios (SNR$>7$) except for M~104 at
43~GHz, where the fringe-fitting failed for many of the scans due to insufficient
SNR. As described below, however, we successfully recovered the solutions for
these scans by referencing the phase/gain solutions derived by J1239--1023 (a few
to several hundreds of mJy so bright enough to detect fringes). Images were
created in DIFMAP software~\citep{shepherd1997} with iterative phase/amplitude
self-calibration. Because M~104 is not so bright, amplitude self-calibration for
this source was performed with relatively long solution intervals
($60\sim300$~minutes, depending on SNR of each data).

\begin{figure}[ttt]
 \centering \includegraphics[angle=0,width=1.01\columnwidth]{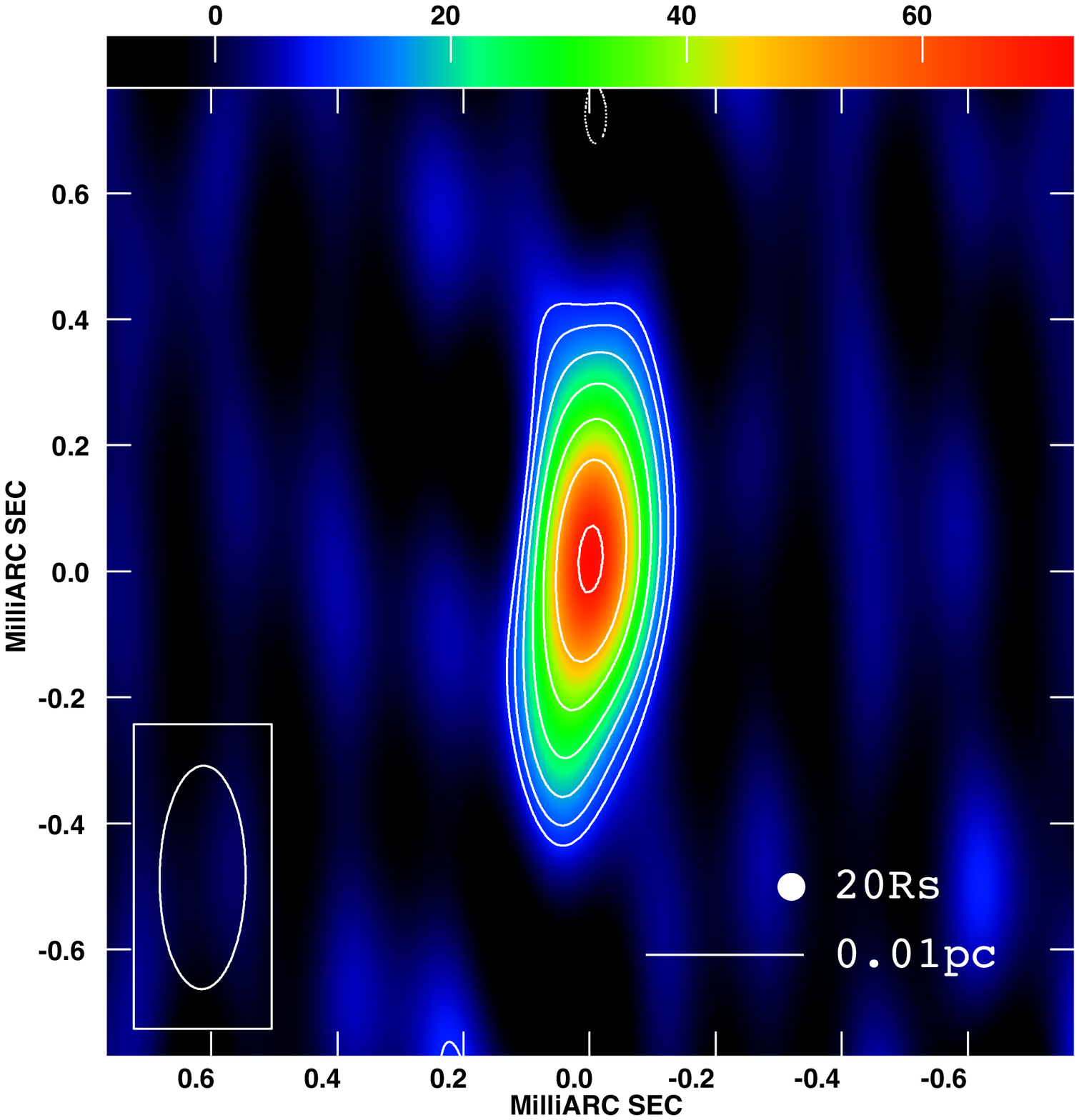} 
  \caption{The highest resolution image of the M~104 nucleus at 43~GHz with
  uniform weighting scheme. The synthsized beam is 0.354~mas $\times$ 0.136~mas at
  $\rm P.A.=-1^{\circ}$ (shown in bottom-left corner of each image). Contours
  start from $-1$, 1, 2, ... times 6~mJy~beam$^{-1}$ (3$\sigma$ image rms) and
  increasing by factors of $2^{1/2}$. The peak flux density is
  73~mJy~beam$^{-1}$. Physical scales in parsec and $R_{\rm s}$ units are
  indicated in the bottom-right corner of the image.} \label{fig:m104q}
\end{figure}

In terms of astrometric investigations for M~104, we performed the
phase-referencing analysis in the following way. After the apriori corrections of
the parallactic angle, ionosphere and instrumental delays, we performed a
fring-fitting only on J1239--1023 with its source model created by the previous
imaging process at each frequency (see Appdendix for the detailed structure this
source). With this process, we derived the solutions for residual
phases/delays/rates, in which the effect of the source structure is already
considered. These solutions were then transferred to the time segments of
M~104. After that, we ran the AIPS task CALIB on J1239--1023 (with the source
model) to derive the corrections of the time-dependent gains for each
antenna. Finally, these gain corrections were transferred to the scans of M~104.
Through these processes, we successfully obtained the phase-referenced images of
M~104, in which the relative position of M~104 with respect to J1239--1023 is
conserved. The astrometry results are shown in section~3.4.
 
Additionally, we found one VLBA archival data of M~104 at 1.4~GHz, which was
observed for a relatively long integration time ($\sim$2.5 hours) with 9 stations
(no Saint Croix). Although this data set is not obtained with astrometry mode and
the epoch is more than 2 years separated from our observations, this is still
useful to understand the nuclear strucrure at lower frequency side. We then
include this data in the present study and analyzed based on the standard data
reduction process.

\begin{table}[ttt]
 \begin{minipage}[t]{1.0\columnwidth}
  \centering \caption{M~104 northern/southern brightness ratio} \smallskip
    \begin{tabular}{cccccc}
    \hline
    \hline
    Frequency & Box in P.A.=$-20^{\circ}$ and $160^{\circ}$& $S_{\rm N}$& $S_{\rm S}$& $R$ \\
    (GHz)  & (radial range from core in mas) &  (mJy)     &  (mJy) &\\
           &  (a)  & (b)   &  (c) & (d)\\
    \hline
    1.430  & 30 -- 50   & 1.2  & 1.1 & 1.1 \\
    2.266  & 10 -- 30   & 2.3  & 2.1 & 1.1 \\
    4.990  & 6 -- 14    & 2.7  & 2.2 & 1.2 \\
    8.416  & 4 -- 10    & 1.8  & 1.4 & 1.3 \\
    15.36  & 2 -- 7     & 2.3  & 1.4 & 1.6 \\
    23.80  & 1.5 -- 3.5 & 2.0  & 0.96 & 2.1 \\
    \hline
    \end{tabular}
  \bigskip
  \medskip
  \end{minipage}
 \label{tab:modelfit} Notes: (a) box region in which the CLEAN components were
 integrated (see Figure~1 for the case at 8.4~GHz): (b) integrated CELAN flux in
 the northern box: (c) integrated CELAN flux in the southern box: (d) brightness
 ratio defined as $S_{\rm N}/S_{\rm S}$. We used source structure models of M~104
 which were created by stacking the two epochs.
\end{table}

\begin{table}[ttt]
 \begin{minipage}[t]{1.0\columnwidth}
  \centering \caption{Geometrical parameters of the modelfitted core} \smallskip
    \begin{tabular}{cccccc}
    \hline
    \hline
    Frequency & $\theta_{\rm maj}$ & $\theta_{\rm min}$/$\theta_{\rm maj}$& P.A. \\
    (GHz)  & (mas) &       &  (deg.) \\
           &  (a)  & (b)   &  (c) \\
    \hline
    1.430  & $11.4\pm5.1$  & 0.09 & $-24\pm$5 \\
    2.266  & $5.22\pm1.8$  & 0.16 & $-24\pm$5 \\
    4.990  & $1.55\pm0.49$ & 0.25 & $-25\pm$5 \\
    8.416  & $0.75\pm0.29$ & 0.06 & $-29\pm$5 \\
    15.36  & $0.37\pm0.12$ & 0.34 & $-31\pm$5 \\
    23.80  & $0.26\pm0.08$ & 0.19 & $-30\pm$5 \\
    43.21  & $0.22\pm0.08$ & 0.21 & $-14\pm$10 \\
    \hline
    \end{tabular}
  \medskip 
  \end{minipage}
 \label{tab:modelfit} Notes: (a) major axis sizes of the derived elliptical
 Gaussians: (b) axial ratios of the derived Gaussian models: (c) position angles
 of the major axes of the Gaussian models.
\end{table}

\section{Results}
\subsection{Images}
In Figure~\ref{fig:images}, we show self-calibrated images of the M~104 nucleus at
all the observed frequencies. Between 2 and 43~GHz, the images shown are made by
stacking the visibility data over the two epochs. We did not find any significant
structural/flux variations at each frequency over the two epochs within the
calibration uncertainties.

Thanks to high qualities of the data set, we have obtained VLBI images for the
M~104 nucleus at unprecedented sensitivities and resolutions. To our knowledge,
this is the first clear VLBI detections and imaging of M~104 at 15, 24 and 43~GHz.
In particular at 43~GHz, where the angular resolution attains $0.35\times0.14$~mas
with uniform weighting scheme, the nuclear structure was imaged on a scale down to
$\sim$60~$R_{\rm s}$ (Figure 2). Along with SgrA*~\citep{shen2005, doeleman2008,
lu2011}, M~87~\citep{junor1999, krichbaum2006, ly2007, hada2011, doeleman2012} and
Cen~A~\citep{tingay1998, horiuchi2006, muller2011}, M~104 is the fourth source in
which such a compact scale was resolved. The obtained image parameters for these
maps are listed in Table~\ref{tab:img_prm}.

In the previous radio observations, only a single compact feature was detected for
the nuclear region both on arcsec~\citep{bruyn1976, hummel1984, bajaja1988,
krause2006} and mas scales~\citep{graham1981, preston1985, shaffer1979}. In the
new experiments presented here, for the first time, we have clearly discovered the
presence of the extended structure elongating from the radio core. This structure
is more remarkable at lower frequencies, and appears to elongate quite
symmetrically toward a northwest (${\rm P.A.}\sim -20^{\circ}$) and a southeast
(${\rm P.A.}\sim 160^{\circ}$) directions. Its entire length is estimated to be
$\sim$120~mas (4.8~pc or $5.4\times 10^{4}~R_{\rm s}$) at 1.4 and $\sim$30~mas
(1.2~pc or $2.7\times 10^5~R_{\rm s}$) at 5~GHz (above 3$\sigma$ level),
respectively. Note that the direction of the extension is similar to that of the
major axis of systhesized beams, but we are confident that this structure is not
an artifact; the extension of this structure is sufficiently larger than each beam
size, and a consistent structure is obtained also at 5~GHz, where the position
angle of the synthesized beam is significantly different from that of the source
elongation. The extended structure is quite smooth and none of knotty features was
found.

\begin{table*}[ttt]
 \begin{minipage}[t]{1.0\textwidth}
  \centering \caption{M~104 spectra for several regions}
   \begin{tabular*}{0.9\textwidth}{@{\extracolsep{\fill}}ccccccccc}
    \hline
    \hline
    Regions & $S_{\rm 1.4}$  & $S_{\rm 2.3}$ & $S_{\rm 5.0}$ & $S_{\rm 8.4}$&
     $S_{\rm 15.2}$ & $S_{\rm 23.8}$ & $S_{\rm 43.2}$ & $\alpha_{\rm ave}$\\
            & (mJy) & (mJy) & (mJy) & (mJy)  & (mJy) & (mJy) & (mJy) &  \\
    \hline
    (1) Total  & $67.3\pm6.7$ & $76.1\pm7.6$ & $90.8\pm9.1$ & $91.0\pm9.1$ & $95.4\pm9.5$ &
			     $92.6\pm9.3$ & $91.2\pm9.1$ & 0.08 \\
    (2) Core  & $59.6\pm6.0$ & $62.1\pm6.2$ & $74.3\pm7.4$ & $80.2\pm8.0$ & $87.1\pm8.7$ &
			     $88.0\pm8.8$ & $91.0\pm9.1$ & 0.14 \\
    (3) Extended & $--$ & $31.9\pm6.4$ & $15.2\pm3.0$ & $8.3\pm1.7$ & $3.8\pm0.8$ &
			   $2.5\pm0.5$ & $< 1.5$ & $- 1.10$ \\
    \hline
   \end{tabular*} \medskip
 \end{minipage}
 \label{tab:addlabel} Notes: see the text for the definition of each region (1),
 (2) and (3). Because of its large beam size (23~mas), reliable measurement of
 $S_{\rm 1.4}$ for the outside of 4-mas-diameter circle was not possible (severe
 blending between the core and the outer emission).
\end{table*}

As frequency increases, the structure becomes progressively compact, and the radio
core increasingly dominates the total radio emission. At 15 and 24~GHz, the
northern side emission tends to be brighter than that of the southern one. We then
estimated a brightness ratio $R$ of the northern/southern emission at each
frequency in the following way; using the source structure model established by
deconvolution process (i.e., a set of CLEAN components), we compared integrated
flux densities of CLEAN components in two rectangular regions, which are at the
same distance from the core having the same size (one is along P.A.=$-20^{\circ}$
and the other is along P.A.=$160^{\circ}$). The results are summarized in
Table~2. $R$ were derived to be $\sim$1.1--1.2 between 1.4 and 5~GHz, whereas
$\sim$1.3--2.1 between 8.4 and 24~GHz. Note that derived value of $R$ at each
frequency varied slightly when we changed the shape of the adopted box, but the
overall trend was basically the same as the case in Table~2 and the northern side
was brighter.  At 43~GHz, almost all of the emission above 3$\sigma$ is confined
within $\sim$1~mas.

\begin{figure}[bbb]
 \centering \includegraphics[angle=0,width=0.85\columnwidth]{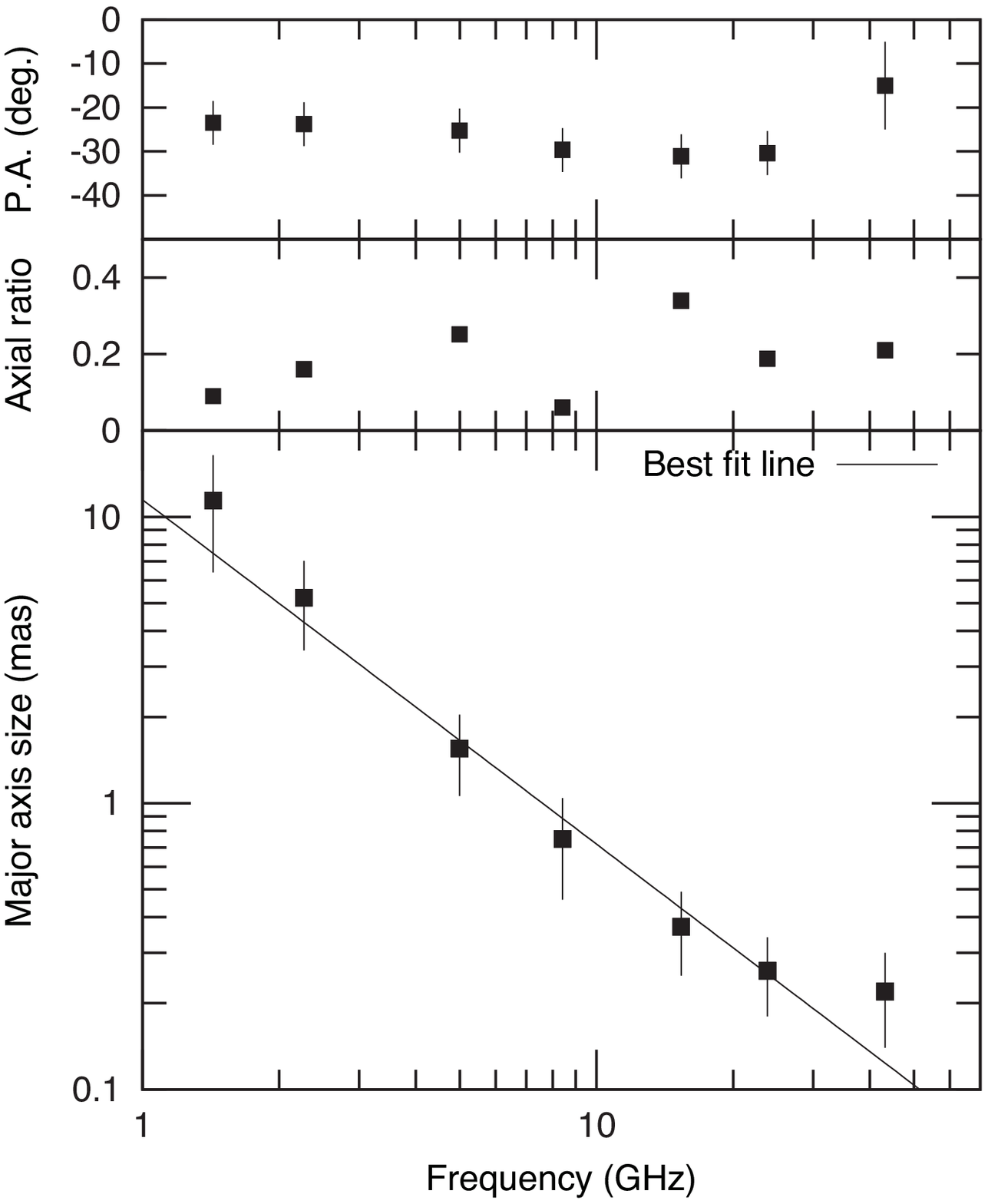}
 \caption{Radio core parameters of the M~104 nucleus as a function of
 frequency. (Top) position angles of major axis of the derived elliptical
 gaussians. (Mid) axial ratios of the derived Gaussian models. (Bottom) major axis
 size of the radio core. The solid line indicates the best-fit solution of the
 frequency dependence of the major axis size $\theta_{\rm maj} \propto
 \nu^{-1.20\pm0.08}$.} \label{fig:core_freq}
\end{figure}

\begin{figure}[ttt]
 \centering \includegraphics[angle=0,width=0.9\columnwidth]{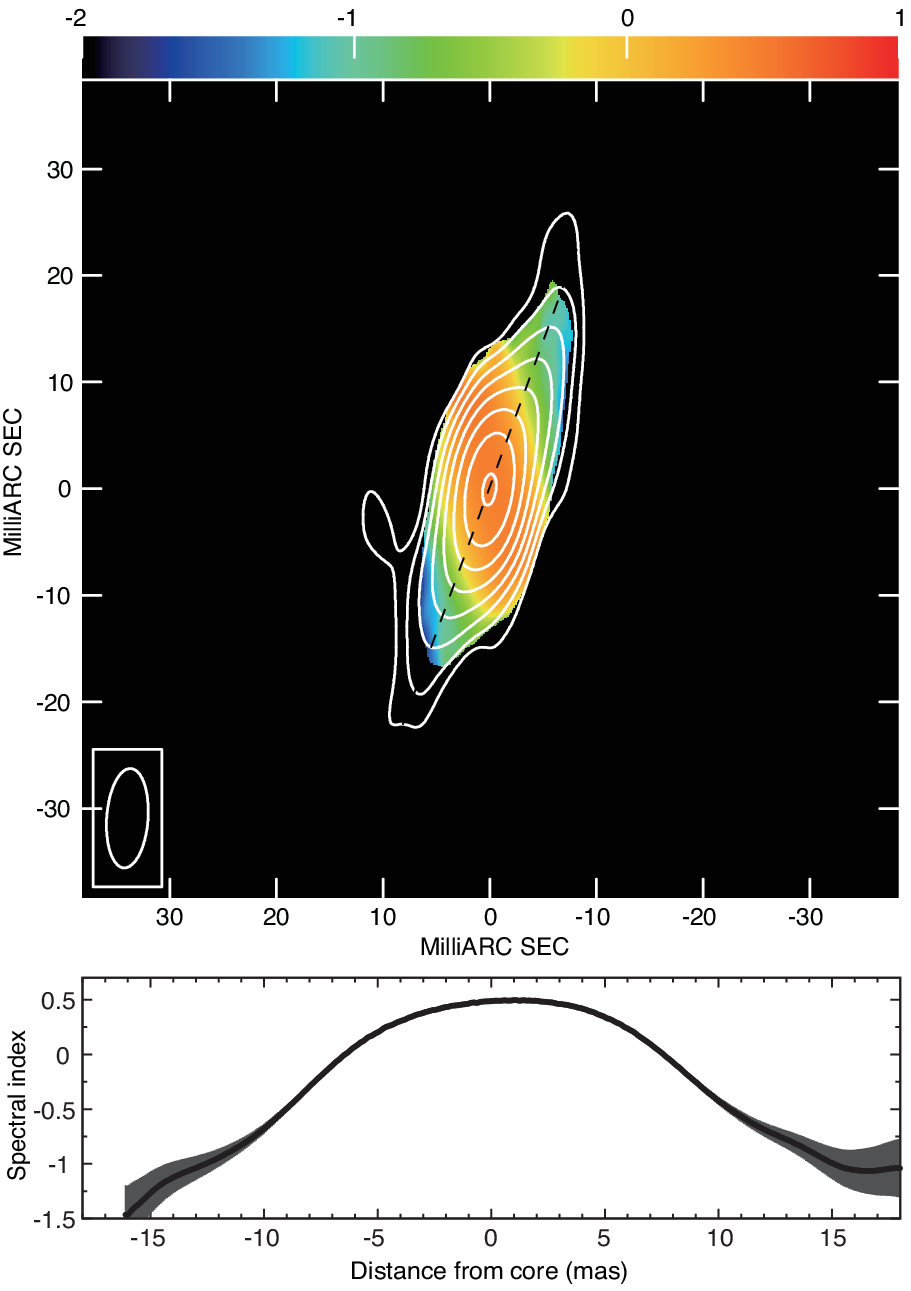}
 \caption{Spectral index distribution map between 2.3 an 5.0~GHz and its slice
 along $P.A.=-20^{\circ}$. The color scale indicates values of spectral index
 $\alpha$, while the overlaid contours indicates the intensity image at
 2.3~GHz. Both of images were commonly convolved with the naturally-weighted
 2.3~GHz beam (shown in the bottom-left corner of the map). The spectral indices
 were then calculated in the region where the flux densities are above 5$\sigma$
 image noise level both at 2.3 and 5.0~GHz. The contours starts from 1, 2,
 4... times $0.4$~mJy~beam$^{-1}$ and increasing by factors of 2. In the slice
 plot, the positive direction of horizontal axis corresponds to the northern
 direction. The gray area in the slice plot indicates an uncertainty of the
 spectral index at each distance, which is estimated using 1$\sigma$ noise level
 of each image.} \label{fig:spix}
\end{figure}

\subsection{Modelfitting}
To quantify the structural properties of the M~104 nucleus, we performed a
modelfitting to the images. As a simple description, here we performed a single
ellitpical Gaussian modelfitting to the core region using the AIPS task JMFIT, and
derived deconvolved parameters of the models. Note that the actual structure is
more complicated as evident from the presence of the extended emission, but this
method is still useful to examine the basic properties of the main emitting
component (i.e., radio core). The obtained values for the geometrical parameters
are summarized in Table~3 and shown in Figure~3 as a function of frequency.

At all the observed frequencies, the derived Gaussian models show highly elongated
shapes in a mean position angle of ${\rm P.A.}\sim -25^{\circ}$ (or
$155^{\circ}$), which smoothly connects to the direction of the larger scale
extension. In terms of the core size, the length of the major axis ($\theta_{\rm
maj}$) is clearly frequency dependent, becoming smaller as frequency increases. We
examined an averaged frequency dependence of the major axis size by fitting a
power-law function to the data between 1.4 and 43~GHz, and obtained the best-fit
solution as $\theta_{\rm maj} \propto \nu^{-1.20\pm0.08}$. On the other hand, the
structure along the minor axis tends to be largely unresolved; the derived lengths
result in typically $\sim$1/5 of the beam size or smaller at each frequency. At
43~GHz, the size of the radio core is derived as $0.22~{\rm (mas)}\times0.08~{\rm
(mas)}$ in ${\rm P.A}=-20^{\circ}$ with a flux density of 91~mJy, corresponding to
$0.009\times0.003$~pc ($1900 \times 600$~AU) or $100\times36~R_{\rm s}$. With this
parameter set at 43~GHz, the brightness temperature of the most compact region is
estimated to be $\sim$$3\times10^{9}$~K.  If the obtained size of the minor axis
is treated as an upper limit, this brightness temperature would be regarded as a
lower limit.

We note that a frequency-dependent size of the radio core is also seen in the M~81
nucleus with a slightly shallower profile \citep[$\propto
\nu^{-0.8}$;][]{bietenholz1996, bietenholz2004, ros2012}. We also remark that no
signs of the scatter-broadening law, which is obviously seen toward SgrA* as
$\theta \propto \nu^{-2} $~\citep{lo1993}, was found in the case of M~104, despite
that a comparable physical scale is observed in gravitational unit.

\begin{figure}[bbb]
 \centering \includegraphics[angle=0,width=1.0\columnwidth]{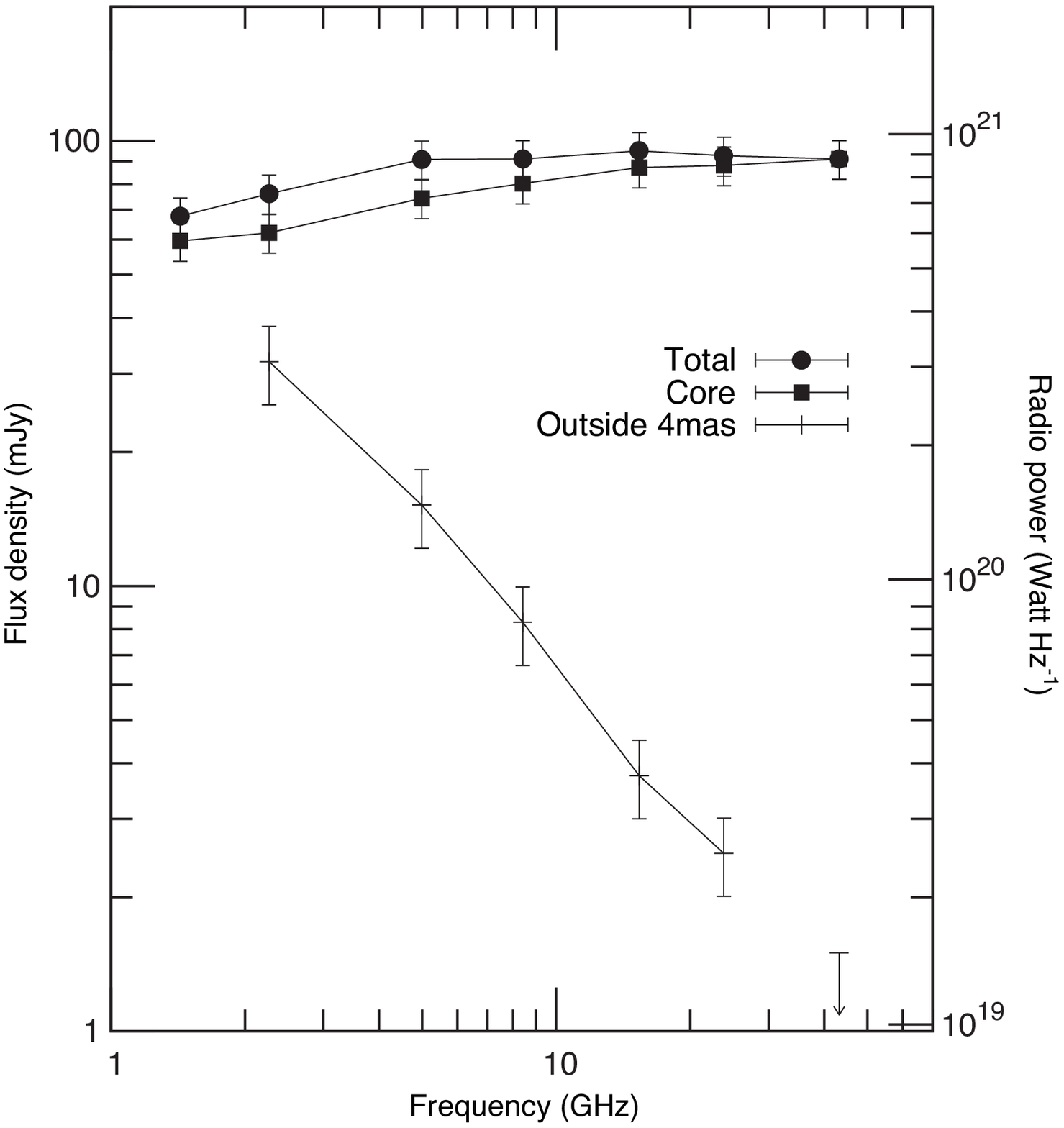}
 \caption{VLBA radio spectra of the M~104 nucleus. The spetra for three different
 regions are plotted: (1) entire CLEANed VLBI flux, (2) radio core flux derived by
 modelfitting in the previous section, (3) integrated CLEANed flux of the extended
 region. For (3), we defined the region as the outside of a 4-mas-diameter circle
 centered on the core.  For the extended region at 43~GHz, we calculated an upper
 limit based on the image rms level.} \label{fig:spec}
\end{figure}

\subsection{Spectra}
Multi-frequency images of M~104 shows the nuclear radio structure to be clearly
frequency dependent, indicating that the spectral property is spatially
varying. We then investigated the spectral properties and the their spatial
distributions. Quasi-simultaneous observations with a wide frequency coverage at
good image qualities allow us to examine these in detail.

In Figure~4, we firstly show a spectral index distribution map between 2 and 5~GHz
and its slice along P.A.$=-20^{\circ}$ in order to qualitatively describe an
overall spectral characteristics. The two images at 2 and 5~GHz are aligned by the
respective phase centers. While a careful alignment is necessary if one seeks for
a more accurate distribution, the frequency-dependent position shift of the radio
core between these frequencies are small compared with the structural extension
(see the next section). Thus Figure~4 gives a reasonable approximation for the
true spectral distribution. As seen in this map, the M~104 nucleus clearly shows a
spatially-varying spectral property. The spatial gradient of the spectral index
distribution is apparent along the source elongation; the spectrum is slightly
inverted near the core and progressively become steep as receding from the core
towards both the northern and southern directions.

\begin{figure}[ttt]
 \centering \includegraphics[angle=0,width=0.95\columnwidth]{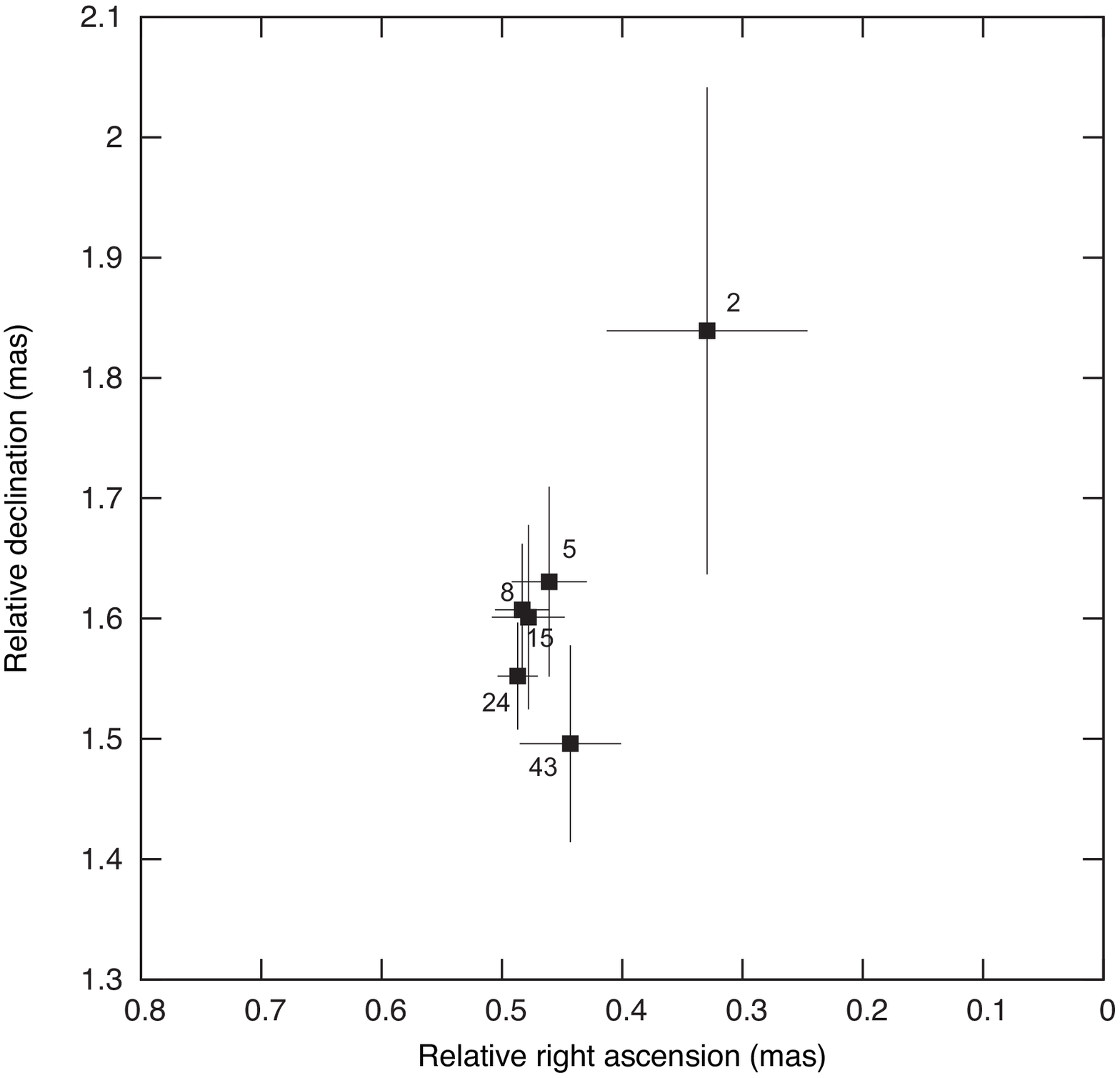}
 \caption{Astrometry of the M~104 core position at each frequency with respect to
 the centroid of an optically-thin feature in J1239--1023. The weighted mean
 position over the two epochs are plotted at each frequency. The number close to
 each data point indicates the observed frequency for each data. The quoted
 position error at each frequency refers to the statistical error budget, which is
 calculated by root-sum-squaring the position uncertainties of the M~104 core
 (phase-referenced image) and the reference feature in J1239--1023
 (self-calibrated image). As the coordinate origin of this plot, the nominal
 phase-tracking center for M~104 ${\rm R.A.}=12^{\rm h}39^{\rm
 m}59.\hspace{-.3em}^{\rm s}4318650$, ${\rm decl.}=
 -11^{\circ}37^{\prime}22.\hspace{-.3em}^{\prime \prime}996380$ (J2000) are used.}
 \label{fig:pos}
\end{figure}

Then, we combined all of the multi-frequency data to clarify more detailed
spectral properties. In Figure~5 and Table~4, we show combined radio spectra
between 1.4 and 43 GHz. Here we plot the spectra for three different regions: (1)
the total CLEANed VLBI flux, (2) the radio core flux derived by the modelfitting
in the previous section, (3) the integrated CLEANed flux for the extended
region. For (3), we defined such a region as the outside of 4-mas-diameter circle
centered on the core, which corresponds to the exterior part of the modelfitted
core at $>$2~GHz. We stress that these spectra are quasi-simultaneously obtained
between 2 and 43~GHz, which is essential to correctly understand the spectral
properties.

Over the observed frequency range, the total emission shows a quite flat spectrum
with an averaged spectral index of $\alpha=0.08\pm 0.03$. However, our
high-resolution observations revealed that such a total spectrum consists of
distinct spectral components. The radio core shows an slightly inverted spectrum
with an averaged spectral index $\alpha=0.14\pm 0.02$, whereas the extended part
shows a steep spectrum with $\alpha =-1.11\pm 0.04$. 

\subsection{Astrometry of the core}
One of the key approaches to specify the origin of nuclear radio emission is to
examine the position shift of the radio core as a function of frequency. This
``core shift'' is expected as a natural consequence of the nuclear opacity effect
if the radio emitting region is dominated by a radio jet~\citep{konigl1981,
lobanov1998, falcke1999, hada2011}. Our quasi-simultaneous, multi-frequency,
phase-referencing observations relative to the calibrator J1239--1023 allow us to
examine this.

In Figure~6, we show the astrometry results of the M~104 core positions. Here we
plot an weighted mean position over the two epochs at each frequency to better
constrain the position. While the jet base structure of J1239--1023 is relatively
complex, we detected a well-isolated, bright optically-thin feature A1 in the
downstream side of this jet (see Appendix on the structure of J1239--1023). So we
measured the relative position of the M~104 core at each frequency with respect to
the intensity centroid of this feature, which is supposed to be stationary on the
sky in terms of frequency~\citep{kovalev2008, sokolovsky2011}. For both of the
M~104 core and the reference feature in J1239--1023, their positions were
registered using a Gaussian model fitting on each image. The quoted position
uncertainties in Figure~6 refer to statistical terms, which are purely caused by
image qualities (phase-referenced images for M~104 core whereas self-calibrated
images for the reference feature in J1239--1023), by calculating the beamsize
devided by the peak-to-noise ratio of each component.

The derived result indicates that the observed core positions tend to move toward a
northern direction as frequency decreases, although the position uncertainties are
rather large because M~104 is a low-declination source. An overall position angle
of the shift is roughly $-10^{\circ}\sim-30^{\circ}$, which is similar to the
direction of the larger-scale northern extension. Mean values of the observed
position shifts are measured to be $\sim$0.2~mas, $\sim$0.3~mas and $\sim$0.4~mas
for 2-5~GHz, 2-24~GHz and 2-43~GHz pairs, respectively.

We should note that, while non-dispersive tropospheric residuals produce a common
systematic position shift at different frequencies (thus such shifts can be
canceled out among frequencies in our quasi-simultaneous observations), dispersive
ionospheric residuals can cause an unwanted frequency-dependent position shift. We
then checked a possible impact of the ionospheric effect by comparing the
astrometry results made with and without the ionospheric correction (AIPS task
TECOR). Phase-referenced images without the correction produced position shifts of
M~104 as large as $\sim$3~mas toward a similar (P.A.$\sim$0$^{\circ}$) northern
direction between 2 and 43~GHz with degrades of image qualities ($\sim$30\%
decrease of the peak-to-noise ratio was seen at 2~GHz). So one cannot completely
exclude the possibility that the observed northward shifts are related to the
ionospheric effect. However, we confirmed that the amount of position differences
between the two methods was well fitted by an inverse-square law with frequency
$\nu^{-2.19\pm 0.20}$, indiating that the ionospheric correction is indeed working
effectively. Using the equation provided in \citet{hada2011}, we estimate a
potential remaining uncertainty due to the ionospheric residuals after the
correction to be $\sim$0.2~mas at 2~GHz, under the observed condition of $\delta
Z=1.\hspace{-.3em}^{\circ}23$, $Z\sim50^{\circ}$ and $\delta I\sim3\times
10^{16}$~m$^{-2}$ where $\delta Z$, $Z$ and $\delta I$ represent the source
separation, source zenith angle and residual total electron content,
respectively. Hence, if the position measurements are still affected by the
ionospheric residuals, the observed $\sim$$0.2$~mas shift for 2-5~GHz pair would
be regarded as an upper limit, and for the pair of 2-24~GHz or 2-43~GHz, a level
of $\sim$0.1-0.2~mas is eventually left for the intrinsic core shift.

\subsection{Variability} 
Previous arcsec-scale observations report that the M~104 nucleus is variable at
radio frequencies at some of different time scales and
amplitudes. \citet{bruyn1976} report a 10$\sim$20\% level variability at 5 and
8~GHz during a period of a few months, whereas \citet{bajaja1988} found a
$\sim$70\% level flux increase at 1.7~GHz between 1971 and 1986. We searched for
any possible variability by comparing the data on March 23 and 30
separately. However, no significant variability was found for any component within
the amplitude calibration accuracy ($\sim$10\%).

The observed VLBI flux densities of $\lesssim 100$~mJy are 20$\sim$40\% lower than
those measured in previous VLBI experiments at 2 and 5~GHz
($\sim$120-140~mJy)~\citep{graham1981, shaffer1979}. This implies the presence of
a long-term variability on mas scale.

\section{Discussion}

Probing the nucleus of M~104 provides clues to understand the physical processes
acting in the vicinity of a very sub-Eddington black hole. Here we discuss the
nuclear structure of M~104 based on the new high-resolution radio results as well
as information at other bands.

\subsection{Evidence for nuclear radio jet}
Because of its highly sub-Eddington luminosity and the absence of Fe K$\alpha$
lines, it is suggested that the standard thin accretion disk is not present in the
nuclear accretion of M~104~\citep{pellegrini2003}. Similarly to other LLAGN, its
accretion state is preferably modeled by RIAF~\citep{dimatteo2001,
yuan2009}. However, a predicted amount of RIAF emission with its Bondi-accretion
rate ($\dot{M}_{\rm Bondi}$) overestimates the observed radio luminosity by a
factor of $\sim$10~\citep{dimatteo2001}; to reconcile the RIAF model, they point
out the necessity of reducing the accretion rate down to a few percent of
$\dot{M}_{\rm Bondi}$ near the inner edge of the accretion flow. Such a inward
decrease of the accretion rate can be realized via outflow/winds~\citep[adiabatic
inflow-outflow solution (ADIOS);][]{bb1999} or
convection~\citep[convection-dominated accretion flow (CDAF);][]{narayan2000},
although recent studies tend to disfavor the latter scenario because the real
accretion flow is convectively stable~\citep{yuan2012, narayan2012}. On the other
hand, RIAF with such a strong mass loss leads to a significant reduction of the
radio-to-X-ray luminosity ratio ($L_{\rm radio}/L_{\rm X}$) from the original
(non-mass loss) RIAF~\citep{quataert1999a}. This situation then leads to a
significant ``underestimate'' of the expected radio emission once the model
attempts to match the observed X-ray luminosity. In the end, these theoretical
studies come to conclude that the presence/dominance of a nuclear radio jet is a
natural solution in order to consistently explain the observed luminosities at
radio/X-ray, and the requirement of mass loss~\citep{dimatteo2001, yuan2009}.
Nonetheless, previous radio observations of M~104 have not found such a jet like
signature so far; only a point-like structure was seen both on
arcsec~\citep{bruyn1976, hummel1984, bajaja1988, krause2006} and
mas~\citep{graham1981, preston1985, shaffer1979} scales, resulting in a major
puzzle for the nuclear structure of M~104.

In this work, our deep VLBA observations have finally obtained compelling evidence
for the nuclear radio jets in the center of M~104. In terms of its morphology, the
pc to sub-pc scale structure is never point-like, but clearly has an extended
component. This extended structure is narrow and collimated with a two-sided
shape, and both sides show steep radio spectra at the observed frequency range. At
the base of the extended structure, the central radio core shows a high brightness
temperature with a flat$\sim$slightly inverted spectrum. The modelfitting shows
the radio core to be highly elongated in the same direction of the outer extended
structure with a clear frequency-dependent size. Moreover, the astrometric
measurement shows a tendency of the core position shift with frequency. While
these characteristics cannot be explained by thermal synchrotron from RIAF, all of
the observed radio properties share the well-known characteristics in more
luminous AGN such as FR I or FR II radio galaxies; the radio emission is created
by nonthermal synchrotron emission from relativistic jets with spatially-varying
nuclear opacities. Note that creating higher brightness temperatures and flatter
spectra of radio core are still possible with RIAF itself if one introduces a
nonthermal electron population in the accretion flow~\citep{yuan2003,
liu2013}. However, such a RIAF-dominated scenario would have still difficulty in
explaining the observed geometrical properties of the radio core; the emission
region of such a hot accretion flow is nearly spherically
symmetric~\citep{narayan1994}, thus does not expect a significant elongation of
the emitting region and a core shift toward a specific direction. Based on a
number of these new evidence, therefore, we argue that the radio emission of the
M~104 nuclues is dominated by the nuclear radio jets emanating from its central
engine over the observed frequencies.

It should be noted that the nuclear radio jets are similarly resolved in the other
nearest cases M~81 (3.6~Mpc) and NGC~4258 (7.2~Mpc) together with an obvious
offset of the radio cores from their putative black hole
positions~\citep{bietenholz2000, bietenholz2004, herrnstein1997, doi2013}. Since
the radio powers of these sources ($P_{\rm 5G}\sim 10^{20.9}, 10^{20.4}$ and
$10^{19.3}$ W~Hz$^{-1}$ for M~104, M~81 and NGC~4258) fall on a roughly
intermediate level among LLAGN population~\citep{nagar2005}, this implies that the
production of jets is a common ability of LLAGN central engines. Regarding more
distant sources, \citet{ulvestad2001} and \citet{anderson2004} have shown in six
prototypical LLAGN (at distances between 15$\sim$40~Mpc with their radio powers
ranging $P_{\rm 5G}\sim 10^{19.5-21}$~W~Hz$^{-1}$, comparable to those of the
above three sources) that their nuclear radio structures are unresolved on mas
scale, but too bright to be explained by RIAF. This situation would be naturally
expected if their jets are highly compact similarly to those seen in M~104, M~81
and NGC~4258. We also note that recent high-sensitivity VLBI studies suggest the
dominance of jet emission in some of the faintest end of LLAGN~\citep[down to
around $10^{19}$~W~Hz$^{-1}$;][]{giroletti2009, bontempi2012}.

\begin{figure}[ttt]
 \centering \includegraphics[angle=0,width=1.0\columnwidth]{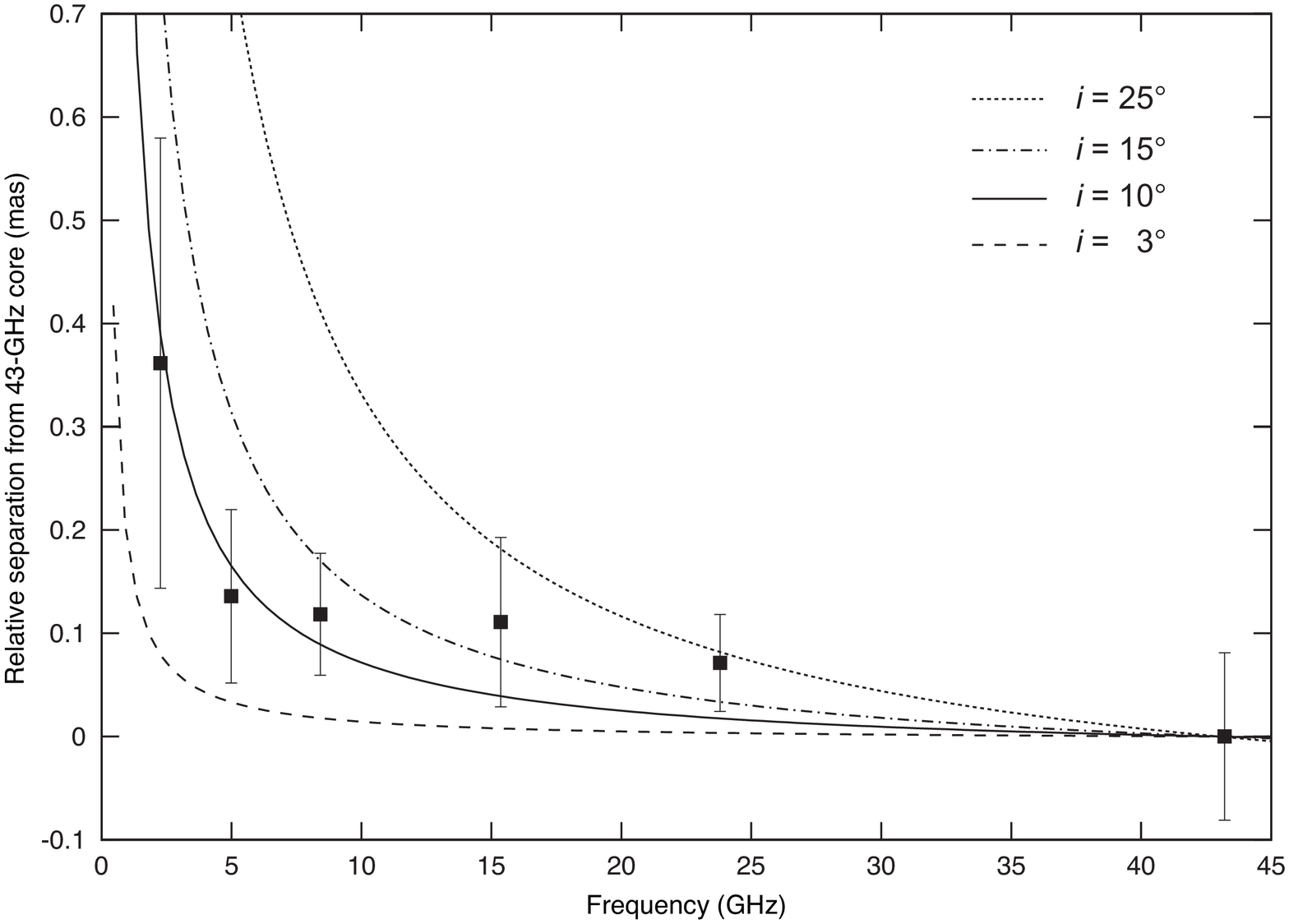}
 \caption{Comparisons of the observed core shift with the compact jet model in
 FB99 for different inclination angles. We plot the observed core separation
 relative to the 43-GHz core position at each frequency. The quoted position
 uncertainty at each frequency is calculated by root-sum-squaring the position
 errors in $x$ and $y$ directions used in Figure~6. In the calculations of the
 FB99 model, we adopt a black hole mass $1.0\times 10^9~M_{\odot}$, a distance
 $d=9.0$~Mpc, a characteristic electron Lorentz factor $\gamma_{e} = 300$, and a
 mean radio core flux density 77~mJy over the observed frequencies. The cases of
 four different inclination angles $i=3^{\circ}$ (long-dash line), $i=10^{\circ}$
 (solid line), $i=15^{\circ}$ (dot-dash line) and $i=25^{\circ}$ (dot line) are
 plotted.}  \label{fig:coreshift}
\end{figure}

\begin{figure}[ttt]
 \centering \includegraphics[angle=0,width=1.0\columnwidth]{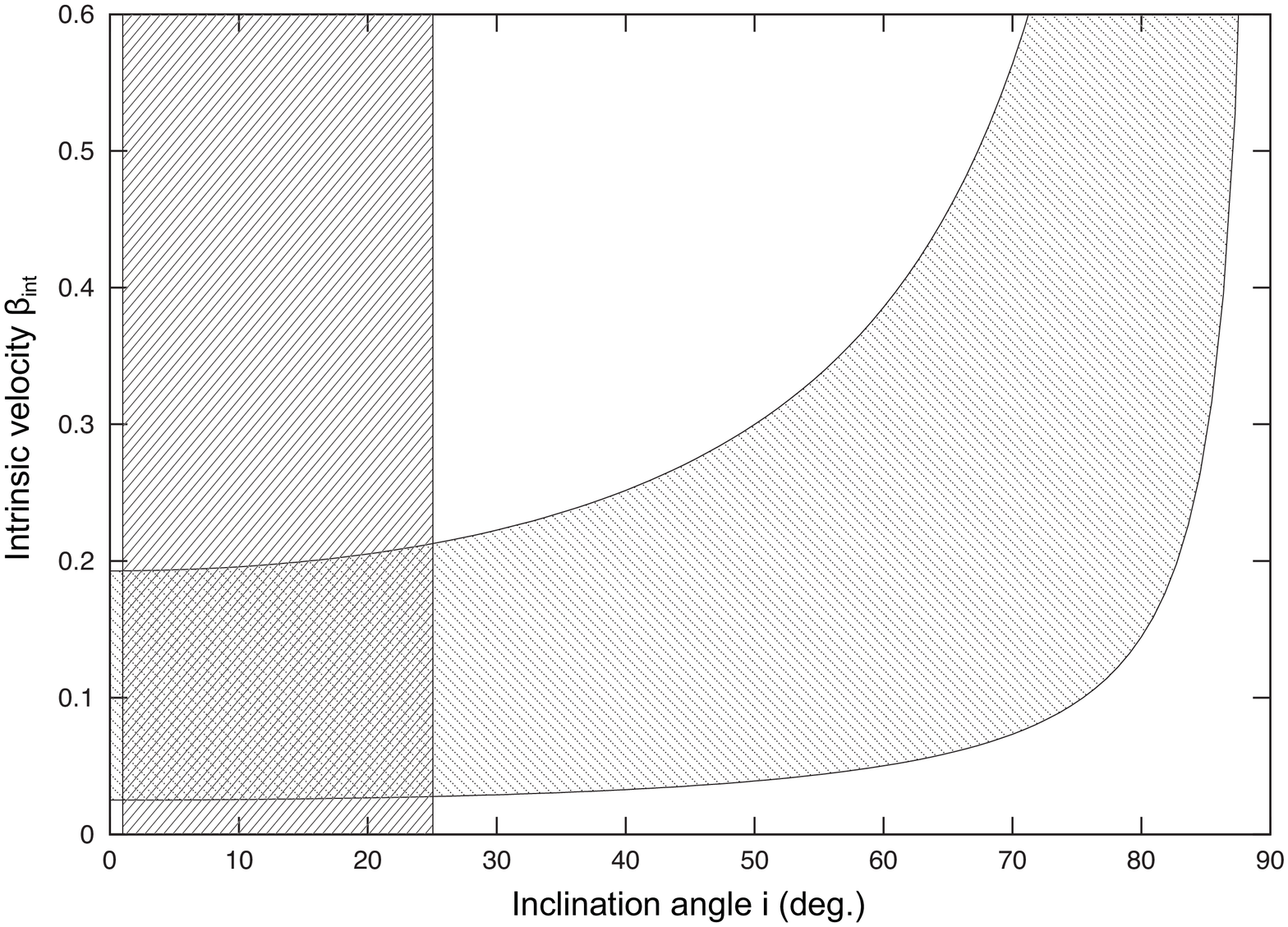}
 \caption{($\beta_{\rm int} - i$) plane for the M~104 jet. The shaded area
 enclosed by the two curved lines indicates the allowed parameter space estimated
 by the brightness ratio between $R=1.1$ and $R=2.1$. The shaded area enclosed by
 the two vertical lines indicates the range of inclination angle constrained by
 the comparison of the observed core shift with the compact jet model by FB99 (see
 Figure~7). The overlapped area of the two regions (in the bottom-left corner of
 the plane) indicates the allowed parameter space for the M~104 jet.}
 \label{fig:coreshift}
\end{figure}

\subsection{Physical properties of the M~104 jet}

We attempt to e1timate fundamental parameters of the M~104 jet based on the
obtained radio results.

First, since the M~104 nucleus shows a two-sided structure, it is important to
determine which is the approaching jet and the other side is receding. Assuming
that the two-sided jet is intrinsically symmetric and any of apparent assymmetry
is caused by Doppler-beaming/de-beaming effects, here we claim that the northern
part is the approaching jet. This is because, while the observed structures at 1.4
and 2.3~GHz are relatively symmetric, at higher frequencies the northern part
tends to become brighter than the southern part with a brightness ratio
$R=1.3\sim2.1$. This argument is also supported by a tendency of the core shift
toward the northern direction as frequency decreases.

Next, by using the brightness ratio of the jet/counter-jet, we estimate an allowed
range of the inclination angle $i$ and the intrinsic velocity $\beta_{\rm
int}=v_{\rm int}/c$. In general, these two parameters are derived simultaneously
by measuring a brightness ratio and an apparent jet velocity. However, the data
used here are not enough to measure the jet motion reliably. Instead, here we
attempt to constrain $i$ by comparing the observed core shift with a theoretical
expectation assuming that the observed radio core at each frequency marks a
synchrotron-self-absorption surface of the approaching jet.

According to the compact jet model proposed by FB99, the amount of the core shift
in LLAGN jets is mainly dependent on observing frequency, characteristic electron
Lorentz factor $\gamma_e$, and inclination angle of the jet. This model has been
applied to several LLAGN such as SgrA*, M~81, NGC~4258 as well as some of galactic
sources, and successfully explain many observational aspects of these sources. In
Figure~7, we show a comparison of the observed core shift (relative to the 43~GHz
core position) with model predictions for several different inclination angles. We
calculated these expected values by combining the equations (8) and (16) in FB99
under the assumption of $\gamma_e=300$, which is suggested to be a typical value
for LLAGN~\citep[similar discussion is presented in][]{anderson2004}.

Because the observed core shift is relatively small, matched models suggest small
inclination angles around $i\sim$3-15$^{\circ}$. For larger $i$, e.g.,
$i=25^{\circ}$, the model still matches the observed core shift between 43~GHz and
15/24~GHz within the uncertainties, but tends to overestimates the values at lower
frequencies. We also checked the impact of $\gamma_{e}$ by changing its value
between 250 and 630, the range of which is implied by the modeling of M~81 and
NGC~4258~(FB99). We confirmed that this uncertainty of $\gamma_{e}$ eases the
allowed range of $i$ to be about 1$^{\circ}$--25$^{\circ}$ in order to reproduce
0.1$\sim$0.7~mas separation between 2 and 43~GHz. In fact, this range of $i$
creates integrated jet luminosities with flat $\sim$ only mildly inverted radio
spectra $\alpha \sim 0-0.10$ in their model ($S_{\nu}\propto \nu^{0.20\xi_2}$
where $\xi_2 \equiv -0.155+1.79i-0.634i^2$; equation (8) in FB99), which is
consistent with the observed integrated spectrum for M~104 ($\alpha \sim 0.08\pm
0.03$). Thus, a range of $1^{\circ}<i<25^{\circ}$ would be a reasonable constraint
on the inclination angle of the M~104 jet.

Then, now we can infer the intrinsic velocity $\beta_{\rm int}$ by combining the
observed brightness ratio and the core shift. Figure~8 shows an allowed range of
$\beta_{\rm int} - i$ plane for the M~104 jet. In this plot, we adopt the
continuous jet model to relate $R$ with ($\beta_{\rm int} - i$) i.e.,
$R=[(1+\beta_{\rm int} \cos\theta)/(1-\beta_{\rm
int}\cos\theta)]^{2-\alpha}$~\citep[e.g.,][]{ghisellini1993}. Here, we adopt
$\alpha=-1.1$ based on the optically-thin part of the observed jet. We found that
the allowed range of the jet velocity results in $0.02 < \beta_{\rm int} < 0.2$
($1.00 < \Gamma < 1.02$) under the ranges of inclination angle
$1^{\circ}<i<25^{\circ}$ and the brightness ratio $1.1<R<2.1$, implying that the
M~104 jet is highly sub-relativistic. Interestingly, this jet speed of M~104 seems
to be quite slower than that suggested for M~81/NGC~4258 based on their VLBI-scale
properties~\citep[roughly $\beta_{\rm int} \gtrsim 0.9$ or $\Gamma \gtrsim$
2-3;][]{doi2013, falcke1999}. Since VLBI observations of M~104 are looking at the
very first stage of jet formation ($100~R_{\rm s}$ scale from the black hole)
compared to those for M~81/NGC~4258 ($10^{3-4}~R_{\rm s}$ or more downstream
region), this discrepancy could be related to some acceleration processes along
the jet. Note that the suggested tendency of a smaller $i$ with a slower
$\beta_{\rm int}$ would become even stronger, if the observed core shift is
affected by the ionospheric residuals.

The derived inclination angle of the M~104 jet has an intriguing implication for
the AGN unification paradigm~\citep{urry1995}. The M~104 nucleus is known to be
one of the representative sources belonging to the ``true type II AGN''; only
narrow optical emission-line components are
detected~\citep[e.g.,][]{nicholson1998} but the amount of obscuring materials
along our line-of-sight is only moderate, as indicated by X-ray
studies~\citep[e.g. $N_{\rm H}=1.8\times 10^{21}$~m$^{-2}$;][]{pellegrini2003} and
also by optical studies~\citep{ho1997b}. If the plane of the circumnuclear torus
is perpendicular to the radio jets \citep[such a condition seems to be actually
realized for nearby AGNs;][]{matsushita2012}, the observed weak obscuration can be
a natural consequence of less intervening material along our line-of-sight (i.e.,
a nearly face-on-view torus). In this case, the M~104 nucleus would actually
belongs to the type I AGN instead of the type II, and thus the apparent
absence/weakness of the broad line components indicates the intrinsic
disappearance/weakness of the broad line region (BLR). Indeed, some theoretical
studies predict the intrinsic disappearance of BLR under certain bolometric
luminosities~\citep[e,g., $L_{\rm bol} < 10^{41.8}(M_{\rm
BH}/10^8M_{\odot})^2$erg~s$^{-1}$;][]{laor2003}, and the M~104's low luminosity
well satisfies this criteria. A face-on view of the M~104 circumnuclear torus is
also suggested by the detection of silicate emission feature in the IR
spectra~\citep{shi2010}.

Finally, we briefly remark a potential importance of the comparison between M~104
and M~87. While these two sources harbor the central black holes with similar
masses, the powerfulness of the radio jets is remarkably different. Such a
distinction may be regulated by the other fundamental parameters of black holes
such as spin or/and accretion rate~\citep{sikora2007}, or physical properties of
the inner part of the accretion flow. In any case, direct vicinity of the black
hole is releveant. M~104 and M~87 are a unique pair for testing this issue because
the black hole vicinity is actually accessibile at a similar horizon-scale
resolution. In particular, the use of ALMA or VLBI at mm-to-submm
regime~\citep{doeleman2012} will be the key, because at such frequencies the
synchrotron emission becomes more transparent to the jet base and most of the
emission comes from the closest part of the central black hole.

\section{Summary}
We investigated the milliarcsecond-scale structure of the M~104 (Sombrero) nucleus
with the dedicated VLBA observations. The followings are new progresses and
findings;

\begin{enumerate}
\item We obtained the VLBI images of the M~104 nucleus, with a drastic improvement
      of sensitivity and resolution, at the seven frequencies between 1.4 and
      43~GHz. At 15, 24 and 43~GHz, this is the first clear VLBI detection and
      imaging for this nucleus. At 43~GHz, we have resolved the nuclear structure
      down to a scale of $\lesssim100~R_{\rm s}$ near the central engine. Along
      with SgrA*, M~87 and Cen~A, M~104 is the the fourth source in which such an
      immediate vicinity of the central black hole was imaged.

\item The innermost region (the radio core) shows a mildly inverted
      ($\alpha\sim0.14$) spectrum between 1.4 and 43~GHz with a brightness
      temperature of $T_{B}\gtrsim 3\times 10^9$~K. The size of the radio core is
      clearly frequency dependent as $\propto \nu^{-1.20}$ with a highly elongated
      shape in a north-south direction. We also obtained a tendency of the core
      position shift toward a northern direction with decreasing frequency. These
      results indicate a nonthermal process near the central enigne, and are
      consistent with synchrotron emission in an optically-thick regime.

\item We revealed the two-sided extended structure on milliarcsecond scale, which
      is extending from the radio core toward northen and southen directions. This
      structure shows a steep spectrum with $\alpha\sim -1.1$. By putting together
      the revealed properties for the radio core and the extended structure, we
      conclude that the central engine of M104 is powering the nuclear radio jets,
      and the jets are overwhelming the radio emission from the underlying RIAF
      over the observed frequencies.

\item Based on the observed brightness ratio, core shift and its comparision with
      a theoretical expectation, we derived the following physical parameters for
      the M~104 jet: (1) the northern side is the approaching jet, while the
      southern part are receding: (2) the inclination angle of the jet is
      relatively close to our line-of-sight, probably less than
      $\sim$$25^{\circ}$: (3) the intrinsic velocity of the jets is highly
      sub-relativistic at a speed of $\beta_{\rm int} \lesssim 0.2$.
\end{enumerate}

\bigskip

\acknowledgments 

 We would like to thank the anonymous referee for his/her careful review and
 valuable comments. The Very Long Baseline Array is operated by the National Radio
 Astronomy Observatory, a facility of the National Science Foundation, operated
 under cooperative agreement by Associated Universities, Inc. This work made use
 of the Swinburne University of Technology software correlator~\citep{deller2011},
 developed as part of the Australian Major National Research Facilities Programme
 and operated under licence. This work was partially supported by KAKENHI
 (24340042). Part of this work was done with the contribution of the Italian
 Ministry of Foreign Affairs and University and Research for the collaboration
 project between Italy and Japan. KH is supported by the Research Fellowship from
 the Japan Society for the Promotion of Science (JSPS).

\appendix
\section{Calibrator J1239--1023}

\begin{figure}[htbp]
 \centering \includegraphics[angle=0,width=0.45\columnwidth]{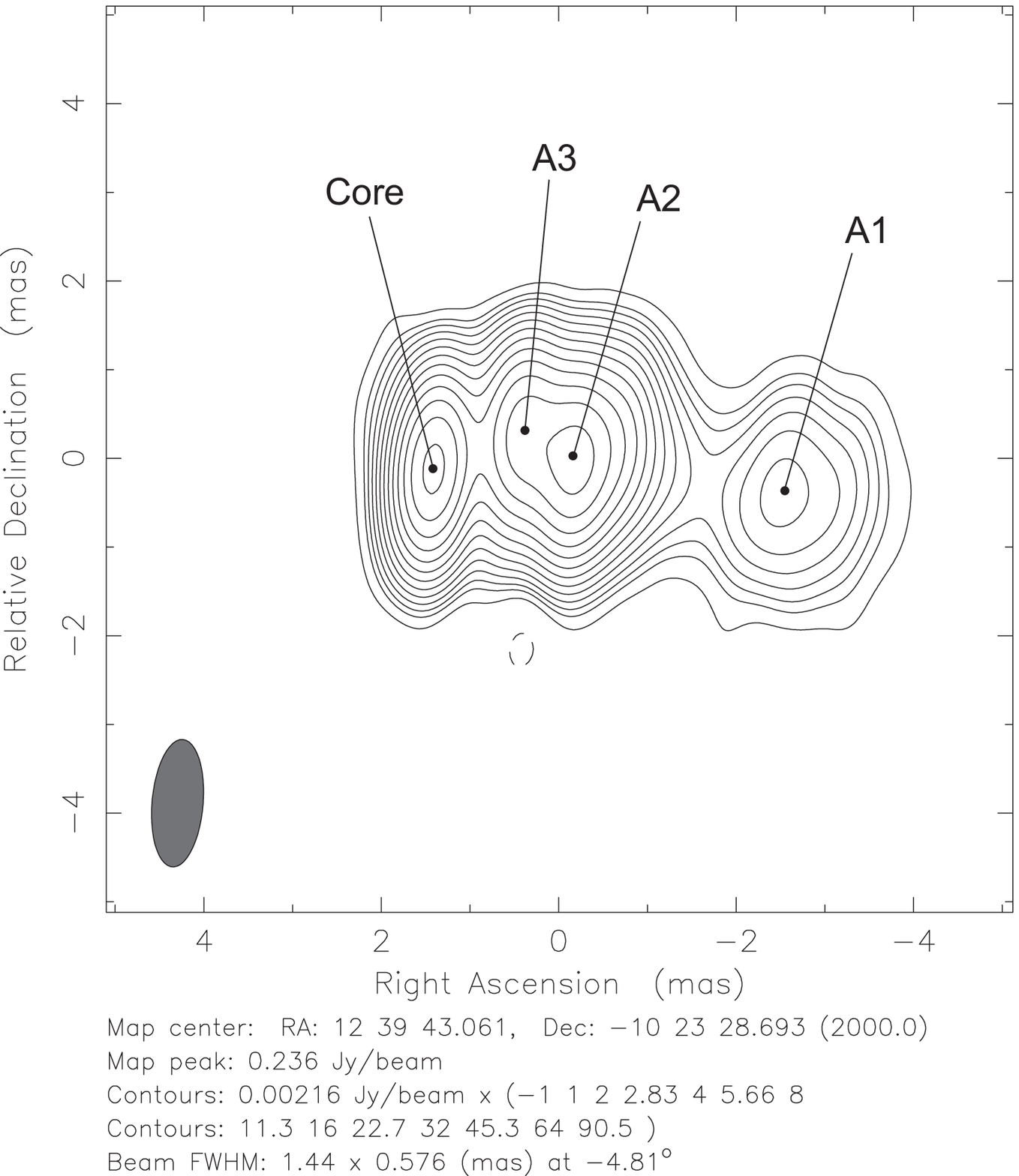}
 \caption{Self-calibrated image of J1239--1023 taken by our VLBA observations at
 15~GHz. Contours start from $-1$, 1, 2, ... times 0.2~mJy/beam (3$\sigma$ image
 rms level) and increasing by factors of $2^{1/2}$. We identified the easternmost
 feature to be the radio core at the jet base becuase of the observed
 optically-thick spectrum at the lower frequency side. We used a bright,
 well-isolated knot A1 as a position reference for the astrometry of M~104. At 15,
 24 and 43~GHz, we also identified additional bright features A2 and A3, both of
 which shows optically-thin spectra at these frequencies. We additionally used
 these features when aligning the images between 24 and 43~GHz.}
 \label{fig:j1239}
\end{figure}

J1239--1023 (1237--101) is a flat spectrum radio quasar~\citep{healey2007} at a
redshift of $z=0.752$~\citep{wisotzki2000}. In Figure~\ref{fig:j1239} we show a
self-calibrated image of J1239--1023 taken by our VLBA observation at 15~GHz. The
source structure is relatively complex with multiple bright components, so in this
case the selection of a reliable reference position becomes a main
concern. Fortunately, we identified a well-isolated, bright feature at $\sim$4~mas
downstream of the jet base. This feature (we term A1) is consistently identified
at all the observed frequencies, and we confirmed its optically-thin nature with a
steep spectrum of $\alpha=-0.8$ between 2 and 43~GHz. We can therefore use this
feature as a reliable position reference.

Supposing that A1 has no frequency-dependent position shifts, we measured relative
positions of the M~104 core with respect to the intensity cenroid of this
feature. At each frequency, we determined the centroid position by fitting a
single Gaussian model to this feature. Position uncertainty of the centroid at
each frequency was estimated by the derived Gaussian size devided by the
peak-to-noise ratio~\citep{fomalont1999}. Because A1 is sufficiently compact
($<1$~mas) and bright (peak-to-noise $>50$) between 2.3 and 24~GHz, the cenroid
positions were determined with an accuracy of $\sim$20~$\mu$as at these
frequencies. Only at 43~GHz, the feature is relatively weak. However, we can still
use the brighter optically-thin components in the upstream region, which are well
resolved in the images at 15, 24 and 43~GHz (A2 and A3 in Figure~9). With the help
of these brighter components, we successfully made the image alignment between 22
and 43~GHz at the same level of position accuracy as that of the lower frequency
data. We confirmed that no significant structural/flux changes or motions were
present over the two epochs.

\end{document}